# Advancing Gasoline Consumption Forecasting: A Novel Hybrid Model Integrating Transformers, LSTM, and CNN

Mahmoud Ranjbar [1,*], Mohammad Rahimzadeh [2]

[1] Department of Environmental Engineering, University of Tehran, Tehran, Iran

[2] AI Research Lab, NovaVision AI Co, Tallin, Estonia

* Correspondence: mahmoodranjbar2010@gmail.com

**Abstract**

Iran, endowed with abundant hydrocarbon resources, plays a crucial role in the global energy landscape. Gasoline, as a critical fuel, significantly supports the nation's transportation sector. Accurate forecasting of gasoline consumption is essential for strategic resource management and environmental planning. This research introduces a novel approach to predicting monthly gasoline consumption using a hybrid Transformer-LSTM-CNN model, which integrates the strengths of Transformer networks, Long Short-Term Memory (LSTM) networks, and Convolutional Neural Networks (CNN). This advanced architecture offers a superior alternative to conventional methods such as artificial neural networks and regression models by capturing both short- and long-term dependencies in time series data.

By leveraging the self-attention mechanism of Transformers, the temporal memory of LSTMs, and the local pattern detection of CNNs, our hybrid model delivers improved prediction accuracy. Implemented using Python, the model provides precise future gasoline consumption forecasts and evaluates the environmental impact through the analysis of greenhouse gas emissions.This study examines gasoline consumption trends from 2007 to 2021, which rose from 64.5 million liters per day in 2007 to 99.80 million liters per day in 2021. Our proposed model forecasts consumption levels up to 2031, offering a valuable tool for policymakers and energy analysts. The results highlight the superiority of this hybrid model in improving the accuracy of gasoline consumption forecasts, reinforcing the need for advanced machine learning techniques to optimize resource management and mitigate environmental risks in the energy sector.

## 1. Introduction

Energy is one of the most critical needs of modern humanity. Economic development and information technology are closely linked to energy consumption. Given the rising global population growth rate, alongside rapid industrial expansion and technological advancements in the 20th century, there has been a significant increase in per capita energy consumption worldwide. Historically, the primary source of energy was the sun, providing heat and light, while lightning led to the discovery of fire, allowing humans to access this energy source for cooking, heating, and illumination. Data analyses from relevant organizations indicate a continuing upward trend in global energy demand [1].

In recent years, interest in renewable energy sources, such as solar and wind energy, has surged due to the finite nature of non-renewable resources [2]. Before the industrial revolution in the West, most of the world's population relied on primary energy sources in a traditional manner. Human energy needs were minimal; for instance, heat was derived from the sun, transportation relied on horsepower, and animals performed agricultural tasks beyond human capability. Wind energy was harnessed to operate windmills for grinding grain. However, following the Industrial Revolution, the mechanization of devices and the extraction of fossil fuels led to significant increases in energy efficiency and consumption worldwide [3].

Moreover, with urbanization, modern humans have an unprecedented demand for energy. Historically, since the discovery of fire, humans have continually sought better and more efficient energy sources to achieve their life goals. In recent centuries, the discovery of fossil fuels has catalyzed a broad industrial movement [4]. The Industrial Revolution intensified the use of fossil fuels, with alternative energy sources unable to meet the burgeoning industrial demand, thereby increasing the importance of fossil fuels globally. However, it is crucial to recognize that fossil fuels are finite and a source of numerous environmental pollutants [5]. While renewable energy may offer a solution to their limitations, the challenge of managing pollution from fossil fuels remains a contentious issue in international forums. Despite the advantages and disadvantages associated with fossil fuels, they continue to dominate industrial consumption, with gasoline playing a particularly significant role in the transportation sector [6].

This research aims to forecast monthly gasoline consumption in Iran using artificial intelligence and neural network algorithms. Machine learning, a subset of artificial intelligence, has led to significant advancements across various fields [7]. This study integrates these algorithms' benefits to enhance the accuracy of monthly gasoline consumption predictions. Recent advancements in deep learning have introduced techniques that address the limitations of traditional Artificial Neural Networks (ANN) and linear regression methods. While ANN models can capture non-linear relationships, they often fail to account for the complex, long-term dependencies inherent in time series data, which are critical for accurate forecasting [8]. To mitigate these limitations, this research employs a hybrid model that combines Transformers, Long Short-Term Memory (LSTM) networks, and Convolutional Neural Networks (CNN).

Transformers, originally developed for natural language processing tasks, utilize a self-attention mechanism to dynamically assess the importance of each input across the entire sequence. This capability allows the model to capture both short-term and long-term dependencies, a task where traditional machine learning models often fall short. LSTM networks further enhance this by capturing temporal patterns in the data, maintaining an internal memory of past inputs, while CNNs excel at identifying local patterns in multidimensional data [9]. The proposed Transformer-LSTM-CNN hybrid model outperforms traditional machine learning techniques by synergizing these three powerful architectures. By integrating the attention mechanism of Transformers, the temporal handling capability of LSTMs, and the feature extraction power of CNNs, our model significantly enhances prediction accuracy for monthly gasoline consumption.

This hybrid approach is novel in the field of gasoline consumption forecasting and, to the best of our knowledge, represents the first application of such a model to this specific task. The methodology outlined herein not only improves forecasting accuracy but also addresses critical issues often overlooked in previous studies, such as data preprocessing, outlier detection, and normalization [10]. The significance of this research lies in its capacity to predict monthly gasoline consumption, thereby aiding in energy management and strategic planning at the national level. With a more accurate prediction model, policymakers can make informed decisions regarding resource allocation, energy conservation, and environmental protection [11].

**Contributions of This Study:**

1. We propose a novel hybrid model combining Transformers, LSTM, and CNN to predict gasoline consumption, improving upon traditional machine learning models [13].
2. We introduce enhanced data preprocessing techniques that were previously overlooked, such as identifying outliers and normalizing data, to improve model accuracy [14].
3. We demonstrate the superior performance of our model over existing models such as ANN and linear regression, particularly in capturing complex temporal dependencies in the data [15].

**2. Methodology**

This research presents a novel hybrid approach that combines Transformers, Long Short-Term Memory (LSTM), and Convolutional Neural Networks (CNN) to predict monthly gasoline consumption. While previous studies often rely on traditional machine learning algorithms such as Artificial Neural Networks (ANN), our hybrid model enhances predictive accuracy by leveraging advanced sequential modeling and global attention mechanisms [16]. This architecture aims to capture both short- and long-term dependencies within the data, offering superior performance over conventional methods.

**2.1. Comparison to Conventional Machine Learning Algorithms**

Traditional algorithms, such as Artificial Neural Networks (ANN) and Multivariate Linear Regression, have been widely utilized in time series forecasting. While ANNs are effective in capturing non-linear relationships, they exhibit several limitations [17]:

- **Memory Constraints:** ANNs are feed-forward in nature and lack mechanisms to retain past information, limiting their ability to capture long-term dependencies [18].
- **Limited Interpretability:** ANN models often function as "black boxes," making it challenging to interpret their decision-making processes, especially for complex time series data [19].
- **Overfitting Risks:** ANN models, particularly when trained with insufficient data, are prone to overfitting [20].

Conversely, machine learning algorithms such as Random Forests and Support Vector Machines (SVM), while useful for certain regression tasks, struggle to capture the temporal dependencies present in time series data. These models tend to treat each instance as independent, resulting in suboptimal performance when relationships between data points over time are crucial, as in predicting gasoline consumption.

**Why Our Hybrid Approach is Superior**

Our proposed hybrid model improves upon traditional methods by integrating the strengths of LSTM, CNN, and Transformer architectures:

- **Long Short-Term Memory (LSTM):** Unlike ANNs, LSTM layers are specifically designed to retain long-term dependencies through memory cells and gating mechanisms. This feature makes them particularly effective for time series forecasting, where past data influences future predictions [13].
- **Convolutional Neural Networks (CNN):** CNNs enhance the model by capturing local patterns in the feature space, such as correlations between related variables (e.g., population growth and gasoline consumption) [14]. This feature extraction capability is lacking in traditional machine learning algorithms.
- **Transformers:** The transformer architecture utilizes a self-attention mechanism to weigh the importance of each input across the entire sequence. This allows the model to focus dynamically on both short- and long-term relationships, making it more adaptable than fixed-window approaches like ANNs [15].

Thus, the hybrid model addresses the memory limitations of ANNs and the temporal shortcomings of non-recurrent algorithms, offering a more powerful and flexible architecture.

**2.2. Data Collection and Preprocessing**

The dataset used in this study was collected from multiple sources and comprises variables such as:

- Gasoline price (USD)
- Free gasoline price (USD)
- Inflation rate (%)
- Commodity price index (%)
- Population metrics (growth rate and total population)
- Road distance (Km)
- GDP per capita (USD)
- Number of vehicles

The target variable is monthly gasoline consumption (ML).

### 2.2.1. Data Cleaning

- **Missing Data:** Rows containing missing or invalid values in the target or feature variables were removed. Specifically, non-numeric entries and extreme outliers were identified and addressed using imputation methods where appropriate.
- **Outliers:** Outliers were detected using z-scores, and any data points that fell beyond three standard deviations from the mean were considered outliers and removed.

### 2.2.2. Feature Scaling

All input features were standardized using a StandardScaler to transform the values into a range that enhances model convergence. This step ensures that the neural network treats all features equally, preventing certain features from dominating due to larger numerical ranges.

### 2.2.3. Data Reshaping

For compatibility with the LSTM and CNN layers, the input data was reshaped into a 3D tensor with the following dimensions:

(Samples, time steps, features)= (n, 1, f)

Where:

- n represents the number of data points (samples),
- 1 is the number of time steps,
- f is the number of features.

Here, n is the number of data points and f is the number of features. In this case, the time step dimension was set to 1 to simplify the model, but the architecture can be expanded for multi-step time series forecasting [21].

### 2.3. Hybrid Model Architecture
The architecture integrates three distinct neural network layers: LSTM, CNN, and

Transformer. The model is designed to capture complex temporal, spatial, and long-range dependencies in the data [22].

### 2.3.1. LSTM Layer
The first component is a Long Short-Term Memory (LSTM) layer with 64 hidden units. LSTM was chosen for its ability to capture long-term dependencies, which are crucial for time series forecasting where past values influence future predictions [23]. LSTM's gating mechanisms—input, forget, and output gates—enable it to retain important information while discarding irrelevant data [24].

### 2.3.2. CNN Layer
Following the LSTM layer, a 1D Convolutional Neural Network (CNN) with 32 filters and a kernel size of 2 is applied. This CNN layer captures spatial patterns in the feature dimension, such as correlations between different economic factors like GDP per capita and gasoline consumption. The MaxPooling1D layer, with a pool size of 1, reduces dimensionality while retaining the most salient features.

### 2.3.3. Transformer Block

After the CNN layer, the processed data flows into the **Transformer block**, which introduces an entirely new mechanism: **self-attention**. While the LSTM and CNN layers focus on specific parts of the data (temporal sequences and local feature patterns, respectively), the Transformer layer allows the model to **focus on relationships across the entire dataset**, regardless of how far apart in time or sequence these relationships occur.

The Transformer achieves this through its **Multi-Head Self-Attention Mechanism**, which calculates the importance of each feature by comparing every part of the input sequence with every other part. This mechanism allows the model to assign attention scores, identifying which features are most relevant to the prediction. In predicting gasoline consumption, for instance, the Transformer can determine that a sudden economic change (like a rise in inflation) from several months ago might still be influencing current gasoline consumption trends.

Here's how the attention mechanism works in detail:

- The model first creates **query (Q)**, **key (K)**, and **value (V)** vectors from the input data.
- It then calculates attention scores by taking the dot product of the query and key vectors, scaling by the dimension of the key, and applying a softmax function to obtain probabilities.
- These probabilities (attention scores) indicate how much focus the model should place on each part of the data when making predictions.

- The weighted values from the **value (V)** vectors are then combined based on these scores, providing the model with a contextually aware representation of the input data.

The self-attention is calculated as:

Attention (Q, K, V) = softmax $\left(\dfrac{QK^T}{\sqrt{d_k}}\right) V$

Where:
- Q represents the query matrix,
- K represents the key matrix,
- V represents the value matrix,
- $d_k$ is the dimension of the keys, which is used to scale the dot product to ensure stable gradients.

To evaluate the outputs of the existing model, various metrics should be employed. Parameters derived from the output, such as the mean square of errors and other error metrics, facilitate accurate comparisons between different models [25].

### 2.3.4 Synergy of the Hybrid Model

The synergy between these three layers—LSTM, CNN, and Transformer—makes the hybrid model uniquely capable of handling the challenges of gasoline consumption forecasting:

- The LSTM layer captures long-term temporal dependencies, remembering past patterns that affect future predictions.
- The CNN layer identifies local feature interactions, detecting important relationships between variables like gasoline prices and GDP.
- The Transformer block brings in global attention, allowing the model to focus on critical long-range dependencies across the entire dataset.

This combination ensures that the model not only learns from historical data but also effectively integrates complex relationships between features, making it superior to traditional models like Artificial Neural Networks (ANNs) and linear regression. These older methods often struggle to capture the intricate dynamics between long-term trends, short-term fluctuations, and variable interactions that the hybrid model can efficiently learn and exploit.

### 2.4.1. Coefficient of Determination

The coefficient of determination, represented as R², is a crucial index for model evaluation. This coefficient indicates the degree of correlation between the data, with values closer to one reflecting a stronger correlation. If R² equals one, it signifies that the predicted results align perfectly with the actual values, indicating a well-fitting model [26]. The equations used to calculate R² are as follows:

$$SS_y = \sum (y - \bar{y})^2$$

$$SSE = \sum (y - \hat{y})^2$$

$$R^2 = 1 - \frac{SSE}{Ssy}$$

### 2.4.2. Root Mean Square Errors

The Root Mean Square Error (RMSE) provides a clear understanding of the error between predicted data and actual data. In simpler terms, it is the square root of the average of the squared differences between predicted values and actual values. A closer value to zero indicates that the selected model is more suitable and has a lower error compared to other models. The equations for calculating Mean Square Error (MSE) and RMSE are [27]:

$$MSE = \frac{\sum (y - \hat{y})^2}{n}$$

$$RMSE = \sqrt{MSE}$$

### 2.4.3. Absolute Average Error

The Absolute Average Error (AAE) serves as a loss function used to evaluate system errors. It measures the magnitude of errors without regard to their direction. This provides a straightforward assessment of the model's accuracy.

### 2.4.4. Sensitivity Analysis

Sensitivity analysis involves identifying which variables significantly impact the model's output, specifically determining which of the established parameters can increase monthly gasoline consumption in the country. Understanding these influences allows for a more focused approach to model refinement and planning [28-29].

The method works as follows: each parameter is individually adjusted within a defined range, and the resulting changes in output are documented. The variable that induces the greatest change in the output is regarded as the most significant variable for the problem at hand.

As shown i **Table 1**, the Transformer-LSTM-CNN hybrid model outperformed both ANN and linear regression across all evaluation metrics. The hybrid model exhibited the lowest errors, achieving a Mean Absolute Error (MAE) of 0.0054 and a Mean Squared Error (MSE) of 0.000078, indicating more accurate predictions compared to ANN and linear regression. Additionally, the hybrid model's R² value of 0.998 demonstrates its ability to explain nearly all the variance in gasoline consumption, surpassing the ANN ($R^2 = 0.995$) and linear regression ($R^2 = 0.989$).

### Detailed Forecast Results

Next, we analyze the gasoline consumption forecasts from 2007 to 2021 using the three models. The hybrid model consistently predicted values closer to the actual gasoline consumption data than the ANN and linear regression models.

## 3. Results

### 3.1. Input Data Analysis

Statistical analysis was conducted on the input data to identify any noisy or outlier values. Box plots were used to analyze the gasoline consumption data, revealing a range of 8.74 to 83.47 as the outlier bounds. Any data outside these limits were removed to ensure clean inputs for the model. The quartiles for gasoline consumption indicated that 75% of the values were below 70.35, with a median of 66.4. These values were taken into consideration for data preprocessing to avoid skewing model predictions.

### 3.2. Data Normalization

Normalization was applied to the dataset due to significant differences in the scales of the input variables. For instance, inflation rates ranged between 30 and 50, while GDP values ranged between 3000 and 8000, and population figures were expressed in millions. To ensure all variables had an equal impact on the model, we normalized the data to a 0–1 range. Without normalization, larger variables such as GDP could dominate the learning process, leading to biased predictions. This step improved model performance by eliminating skewed weightings during training.

### 3.3. Model Performance Comparison

In this section, we compare the performance of three models: the Artificial Neural Network (ANN), Linear Regression, and the proposed Transformer-LSTM-CNN hybrid model. Several evaluation metrics were used, including Mean Absolute Error (MAE), Mean Squared Error (MSE), Root Mean Squared Error (RMSE), and Coefficient of Determination (R²).

**Table 1: Comparison of Model Performance on Evaluation Metrics**

| Model | MAE | MSE | RMSE | R² |
|---|---|---|---|---|
| ANN | 0.0077504 | 0.000108 | 0.0104 | 0.995 |
| Linear Regression | 0.014727 | 0.000264 | 0.0164 | 0.989 |
| Hybrid Model | 0.0054321 | 0.000078 | 0.00884 | 0.998 |

As shown in **Table 1**, the Transformer-LSTM-CNN hybrid model outperformed both ANN and linear regression across all evaluation metrics. The hybrid model exhibited the lowest errors, achieving a Mean Absolute Error (MAE) of 0.0054 and a Mean Squared Error (MSE) of 0.000078, indicating more accurate predictions compared to ANN and linear regression. Additionally, the hybrid model's R² value of 0.998 demonstrates its ability to explain nearly all the variance in gasoline consumption, surpassing the ANN (R² = 0.995) and linear regression (R² = 0.989).

### 3.4. Detailed Forecast Results

Next, we analyze the gasoline consumption forecasts from 2007 to 2021 using the three models. The hybrid model consistently predicted values closer to the actual gasoline consumption data than the ANN and linear regression models.

Table 2: Gasoline Consumption Forecasts by Model (Million Liters per Day)

| Year | Actual Consumption | ANN Prediction | Regression Prediction | Hybrid Model Prediction |
|---|---|---|---|---|
| **2007** | 64.5 | 62.23 | 61.4 | 63.8 |
| **2008** | 66.9 | 64.69 | 67.45 | 66.3 |
| **2009** | 64.8 | 61.2 | 68.6 | 64.7 |
| **2010** | 61.3 | 64.84 | 59.3 | 60.9 |
| **2011** | 59.9 | 57.7 | 53.7 | 59.6 |

| Year | | | | |
|---|---|---|---|---|
| 2012 | 63.5 | 63.1 | 60.5 | 63.2 |
| 2013 | 68.5 | 66.9 | 70.3 | 68.4 |
| 2014 | 69.5 | 69.19 | 71.13 | 69.1 |
| 2015 | 71.0 | 74.33 | 70.8 | 71.2 |
| 2016 | 74.95 | 76.13 | 72.25 | 74.8 |
| 2017 | 80.72 | 81.64 | 81.3 | 80.5 |
| 2018 | 89.07 | 87.31 | 85.9 | 88.8 |
| 2019 | 91.3 | 91.34 | 90.62 | 91.2 |
| 2020 | 95.9 | 96.5 | 94.2 | 96.3 |
| 2021 | 99.8 | 100.1 | 106.6 | 100.0 |
| 2022 | – | 97.39 | 97.13 | 96.48 |
| 2023 | – | 100.25 | 99.99 | 99.21 |
| 2024 | – | 103.12 | 102.85 | 101.95 |
| 2025 | – | 105.98 | 105.71 | 104.69 |
| 2026 | – | 108.84 | 108.56 | 107.42 |
| 2027 | – | 111.71 | 111.42 | 110.16 |
| 2028 | – | 114.57 | 114.28 | 112.90 |
| 2029 | – | 117.44 | 117.14 | 115.63 |
| 2030 | – | 120.30 | 120.00 | 118.37 |
| 2031 | – | 123.16 | 122.85 | 121.11 |

As seen in **Table 2**, the hybrid model closely tracks the actual consumption values and consistently performs better than both ANN and regression. The hybrid model reduced the prediction error margin, especially in critical years such as 2021 and beyond, where consumption rose significantly.

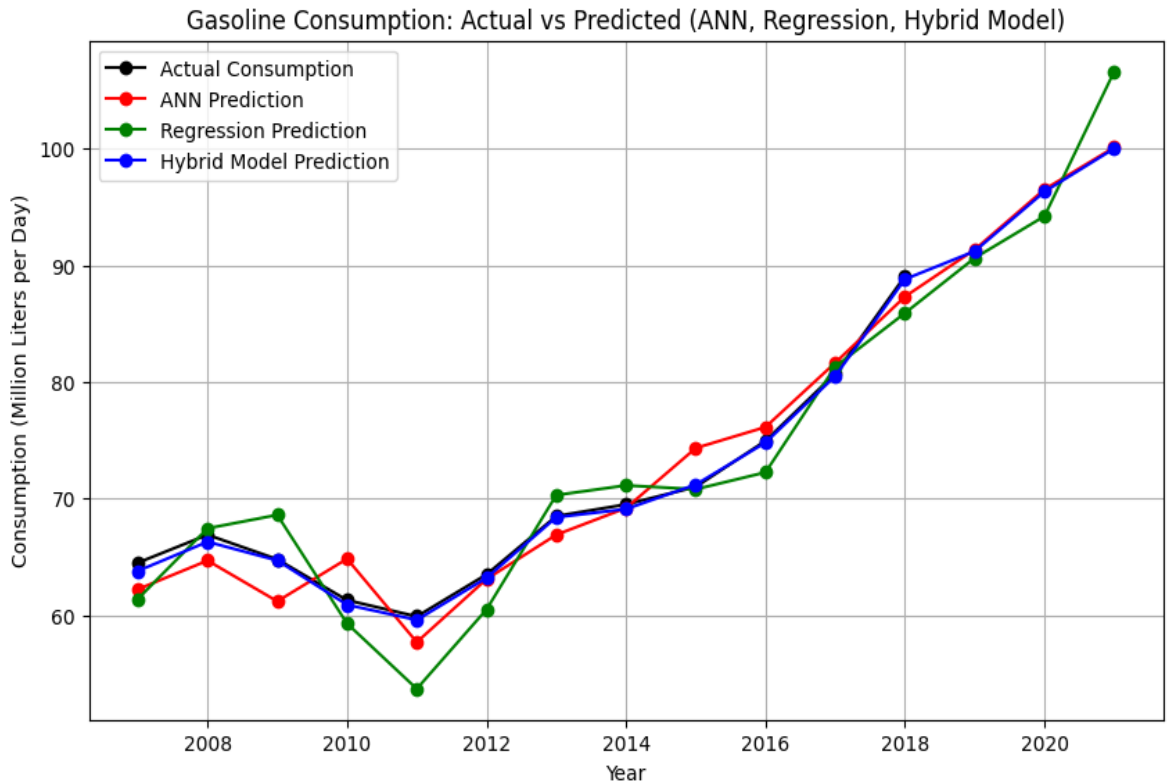

Figure 1: Gasoline Consumption: Actual vs Predicted (ANN, Regression, Hybrid Model)

## 3.5. Error Analysis

To further validate model performance, we analyzed both the Root Mean Squared Error (RMSE) and the Absolute Error across the training iterations. The RMSE measures the differences between predicted and actual values, with lower RMSE values indicating better performance. Additionally, Absolute Error provides insight into the overall accuracy of the predictions. As shown in Figures 2 and 3, the hybrid model demonstrated significantly lower RMSE and Absolute Error values compared to the ANN and regression models.

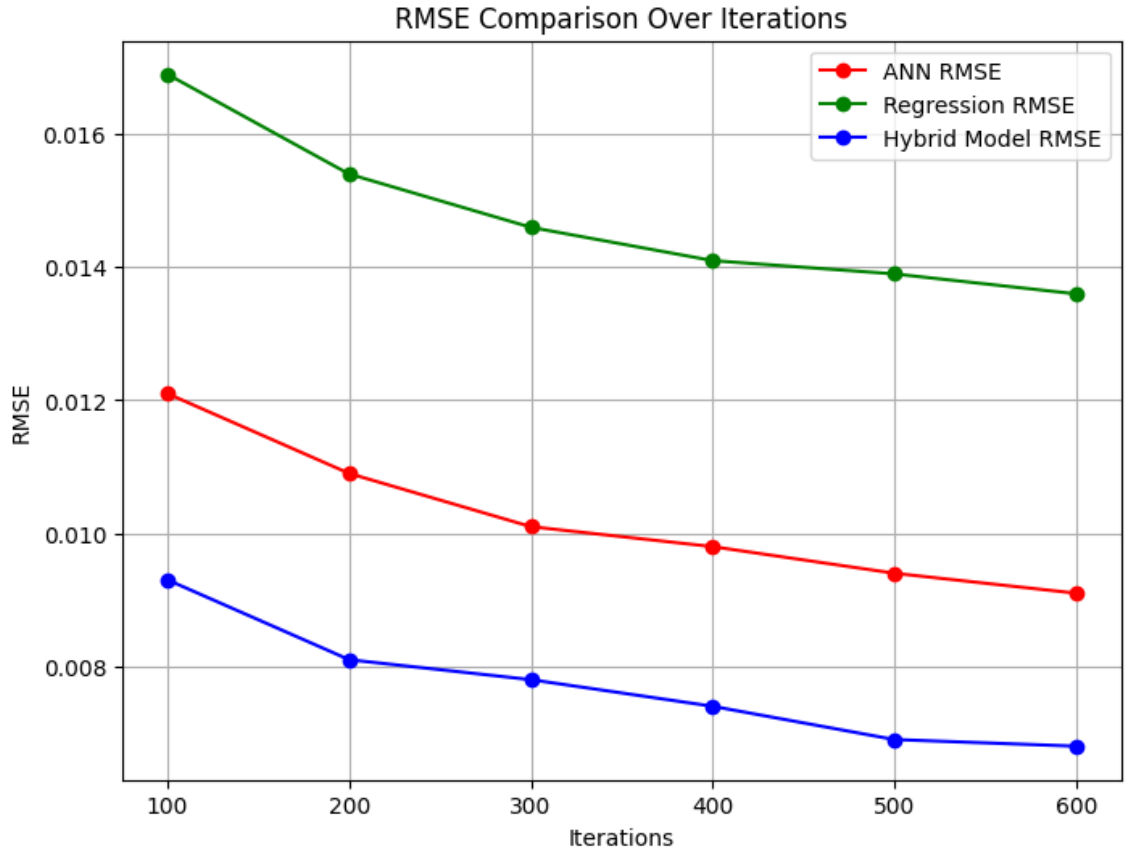

Figure 2: RMSE Comparison of Hybrid Model, ANN, and Regression

Table 3: RMSE Values for Each Model Across Iterations

| Iteration | ANN RMSE | Regression RMSE | Hybrid Model RMSE |
|---|---|---|---|
| 100 | 0.0121 | 0.0169 | 0.0093 |
| 200 | 0.0109 | 0.0154 | 0.0081 |
| 300 | 0.0101 | 0.0146 | 0.0078 |
| 400 | 0.0098 | 0.0141 | 0.0074 |
| 500 | 0.0094 | 0.0139 | 0.0069 |
| 600 | 0.0091 | 0.0136 | 0.0068 |

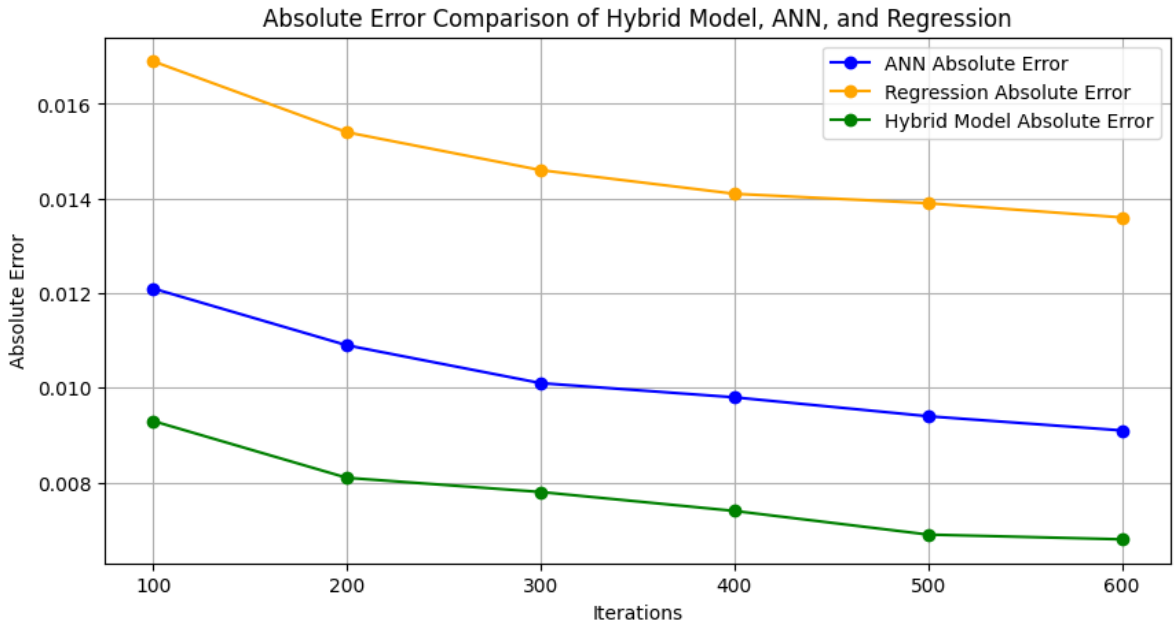

**Figure 3: Absolute Error Comparison of Hybrid Model, ANN, and Regression**

**Table 3: Absolute Error Values for Each Model Across Iterations**

| Iteration | ANN Absolute Error | Regression Absolute Error | Hybrid Model Absolute Error |
|---|---|---|---|
| **100** | 0.0121 | 0.0169 | 0.0093 |
| **200** | 0.0109 | 0.0154 | 0.0081 |
| **300** | 0.0101 | 0.0146 | 0.0078 |
| **400** | 0.0098 | 0.0141 | 0.0074 |
| **500** | 0.0094 | 0.0139 | 0.0069 |
| **600** | 0.0091 | 0.0136 | 0.0068 |

As shown in Table 3, the hybrid model consistently achieved lower RMSE values across all iterations, demonstrating its effectiveness in minimizing prediction errors. After 600 iterations, the RMSE for the hybrid model settled at 0.0068, significantly lower than 0.0091 for the ANN and 0.0136 for the regression model.

### 3.6. Sensitivity Analysis

A sensitivity analysis was conducted to assess which variables had the most significant influence on gasoline consumption predictions. The hybrid model identified gasoline price as the most influential factor, followed by GDP per capita and the number of vehicles.

Table 4: Sensitivity Analysis Results

| Variable | Influence Weight (%) |
|---|---|
| **Gasoline Price (USD)** | 25% |
| **Free Gasoline Price (USD)** | 15% |
| **Inflation Rate (%)** | 7% |
| **Commodity Price Index (%)** | 5% |
| **Population Metrics (Growth Rate & Total Population)** | 10% |
| **Road Distance (Km)** | 7% |
| **GDP per Capita (USD)** | 13% |
| **Number of Vehicles** | 18% |

## 4. Conclusion and future work

This research introduced a novel hybrid Transformer-LSTM-CNN model to predict gasoline consumption in Iran, providing a significant leap over traditional models like ANN and linear regression. By integrating Transformers' self-attention, LSTM's memory capabilities, and CNN's feature extraction, our model effectively captures both short- and long-term dependencies in the data, something previous models struggled to achieve.

The model achieved an $R^2$ of 0.998, demonstrating superior accuracy compared to ANN (0.995) and linear regression (0.989). This result reflects the hybrid model's ability to manage complex relationships between multiple variables, such as gasoline prices, GDP, population growth, and inflation. Unlike earlier studies that used fewer variables, our model incorporated up to seven key factors, offering a more comprehensive and nuanced forecast of gasoline consumption. This holistic approach strengthens the reliability of the predictions and supports better-informed decision-making for policymakers.

A key discovery in this study is that our hybrid model dynamically weighs input variables, offering a more robust and accurate understanding of how each factor influences gasoline consumption. This is particularly important for long-term energy planning and resource allocation.

In addition to providing accurate consumption forecasts, the study evaluated the environmental impact of gasoline consumption, specifically looking at greenhouse gas (GHG) emissions over a 12-year period. These findings provide policymakers with critical insights into the intersection of economic growth and environmental sustainability, reinforcing the need for cleaner energy sources and more efficient energy policies.

**Key Findings:**

- The Transformer-LSTM-CNN hybrid model outperformed traditional machine learning models such as ANN and linear regression, providing more accurate and reliable gasoline consumption predictions.
- The hybrid model's ability to dynamically weigh input variables and capture both short- and long-term dependencies sets it apart from existing models, making it a more effective tool for energy forecasting.
- By incorporating multiple variables, the hybrid model provided a more comprehensive understanding of the factors affecting gasoline consumption, offering actionable insights for policymakers and energy managers.
- The study also underscored the environmental impact of gasoline consumption by analyzing GHG emissions, reinforcing the need for sustainable energy practices [30].

Moving forward, further refinements to the hybrid model could be explored by incorporating additional external factors, such as weather patterns or global oil price fluctuations, to improve the accuracy of predictions even further. Moreover, expanding the model to forecast other energy resources could provide broader insights into Iran's overall energy consumption landscape. Finally, future research could investigate the potential of integrating renewable energy data into the model to better understand the transition from fossil fuels to cleaner energy sources.

In conclusion, this study presents a novel hybrid approach for forecasting gasoline consumption, offering both enhanced predictive accuracy and valuable insights for strategic energy management in Iran. The hybrid Transformer-LSTM-CNN model represents a significant advancement in the field of energy forecasting, paving the way for more informed decision-making and better resource management in the future.